%% file: submission.tex
\documentclass[runningheads]{llncs}

\usepackage{eccv}

\usepackage{eccvabbrv}

\usepackage{graphicx}
\usepackage{booktabs}

\usepackage{multirow}
\usepackage{makecell}
\usepackage{wrapfig}
\usepackage{xcolor} 
\usepackage{pifont}
\usepackage[accsupp]{axessibility}  

\usepackage{booktabs}
\usepackage{multicol}
\usepackage{colortbl}
\usepackage{multirow}
\usepackage{amssymb}
\usepackage{comment}
\usepackage[normalem]{ulem}
\usepackage{xcolor}

\definecolor{cvprblue}{rgb}{0.21,0.49,0.74}
\usepackage[pagebackref,breaklinks,colorlinks,citecolor=cvprblue]{hyperref}

\title{DGR-MIL: Exploring Diverse Global Representation in Multiple Instance Learning for Whole Slide Image Classification}

\titlerunning{DGR-MIL}

\footnotetext{These authors contributed equally to this paper.}
\footnotetext{Corresponding author}

\author{Wenhui Zhu\inst{1* \dagger} \and
Xiwen Chen\inst{2*} \and
Peijie Qiu\inst{3*}  \and \\
Aristeidis Sotiras\inst{3} \and Abolfazl Razi\inst{2} \and Yalin Wang\inst{1} \\
}
\authorrunning{Zhu. Wenhui et al.}
\institute{ Arizona State University, AZ, USA \\ {\tt\small \{wzhu59,ylwang\}@asu.edu} \and
Clemson University, SC, USA \\ {\tt\small xiwenc@g.clemson.edu, arazi@clemson.edu} \and
Washington University in St. Louis, MO, USA \\
{\tt\small \{peijie.qiu,aristeidis.sotiras\}@wustl.edu }
}

\begin{document}
\maketitle

\begin{abstract}
Multiple instance learning (MIL) stands as a powerful approach in weakly supervised learning, regularly employed in histological whole slide image (WSI) classification for detecting tumorous lesions. However, existing mainstream MIL methods focus on modeling correlation between instances while overlooking the inherent diversity among instances. However, few MIL methods have aimed at diversity modeling, which empirically show inferior performance but with a high computational cost.
To bridge this gap, we propose a novel MIL aggregation method based on diverse global representation (DGR-MIL), by modeling diversity among instances through a set of global vectors that serve as a summary of all instances. First, we turn the instance correlation into the similarity between instance embeddings and the predefined global vectors through a cross-attention mechanism. This stems from the fact that similar instance embeddings typically would result in a higher correlation with a certain global vector. Second, we propose two mechanisms to enforce the diversity among the global vectors to be more descriptive of the entire bag: (i) positive instance alignment and (ii) a novel, efficient, and theoretically guaranteed diversification learning paradigm. Specifically, the positive instance alignment module encourages the global vectors to align with the center of positive instances (e.g., instances containing tumors in WSI). To further diversify the global representations, we propose a novel diversification learning paradigm leveraging the determinantal point process. The proposed model outperforms the state-of-the-art MIL aggregation models by a substantial margin on the CAMELYON-16 and the TCGA-lung cancer datasets. The code is available at \url{https://github.com/ChongQingNoSubway/DGR-MIL}.
\begin{figure}[!t]
    \centering
    \includegraphics[width=1.0\columnwidth]{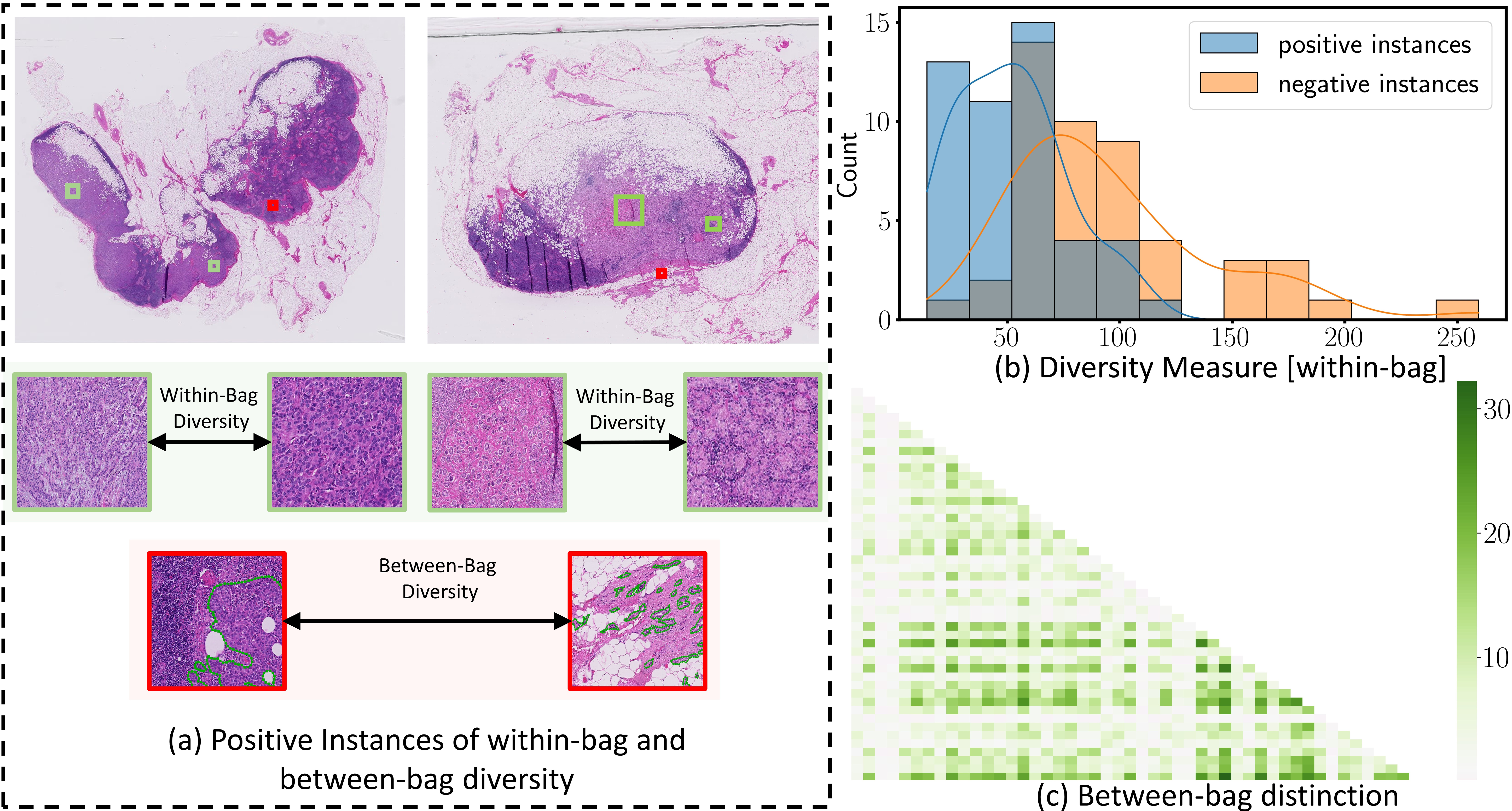}
    \caption{(a) Examples of positive instances of with-bag and between-bag diversities measured by rate-distortion theory. (b) Histogram of the diversity measure within positive bags on the CAMELYON16 dataset. (c) The between-bag distinction measures the pair-wise similarity between bags.}  
    \vspace{-0.4cm}
    \label{fig:div_vis}
\end{figure}
  \keywords{Weakly-supervised learning, Multiple Instance Learning \and Histological Whole Slide Image \and Transformer}
\end{abstract}

\section{Introduction}
Histological whole slide images (WSIs) are commonly used to diagnose a variety of cancers, e.g., breast cancer, lung cancer, etc.~\cite{cancer}. However, the gigapixel resolution of WSIs hinders the direct translation of classic deep learning methods into WSI applications mainly due to computational intractability~\cite{wsi1,wsi2,wsi3,clam-sb}. Therefore, the analysis of WSIs typically starts with cropping images into small patches and then performing analysis on a per-patch basis. In addition, the absence of labor-intensive pixel/patch-level annotations poses a significant challenge for the precise localization of targets of interest (e.g., tumors in WSIs) in a fully supervised setting. As a result, Multiple Instance Learning (MIL), a weakly supervised method, is commonly employed in WSI analyses by treating an entire WSI as a bag and the cropped patches as instances. \\
\indent The prevailing MIL models in analyzing WSIs have been built upon the attention-based MIL (AB-MIL) framework ~\cite{ilse2018attention} since its introduction. However, the standard AB-MIL treats each instance independently and does not take the correlations between instances into account. Although many of its follow-ups address this challenge by a variety of means~\cite{li2021dual,shao2021transmil,zhang2022dtfd,lowrankmil}, they mainly focus on modeling the correlation between instances by assigning high correlations to instances from the same category (e.g., tumor instances). However, even instances from the same category exhibit variations in phenotype, size, as well as spatial diversity marked by immune infiltration across different patients~\cite{clinicdiversity,clinicdiversity2,clinicdiversity3}. For example, negative instances close to the tumor boundaries typically resemble positive instances while appearing differently compared to the other negative instances~\cite{similaritysupport}. As a result, instances belonging to the same category may not be assigned high correlations; similarly, instances from different categories could also receive high correlations. This spurious correlation between instances is prone to trap the MIL model by incorrectly aggregating them when making predictions. Formally, we quantify the diversity of instances between and within bags in WSIs by leveraging the rate-distortion theory~\cite{cover1999elements,yu2020learning,chen2023rd}, where a higher rate indicates a less compressible but more diverse collection of samples (see details of computing the diversity measure in Appendix A). As consistent with findings in pathology, we observe that both positive and negative instances in WSIs exhibit between-bag and within-bag diversity (refer to Fig.  \ref{fig:div_vis}). Based on this fact, we argue that the diversity of instances is important in designing MIL models. Before that, clustering/prototype-based MIL methods tried to solve the diversity by utilizing attention scores as pseudo labels to provide instance-level supervision~\cite{tpmil,scmil}. This introduces a chicken-and-egg issue. The effectiveness of pseudo-labels relies on successful MIL classification pooling, which in turn depends on precise attention localization. Especially when patch representations are inferior or MIL initially guided by poor pseudo label, leading to even misleading localization and unstable optimization~\cite{iterself,PDL}. Among them, PMIL presents an alternative method to avoiding noise attention~\cite{pmil}, initially selecting prototypes through clustering, followed by modeling diversity via prototype and patch representation. However, the design of the multi-stage framework empirically leads to suboptimal learning outcomes, and the restricted number of prototypes, due to high computational burden, results in diminished diversity. \\
\indent To this end, we propose to jointly model this diversity through a set of learnable global vectors. The learned global vectors summarize diverse instances of interest (e.g., tumors in WSIs). As a result, the diversity between instances can be implicitly modeled by computing the correlation between instance embeddings and the global vectors through a cross-attention mechanism. To enhance the ability of the global vectors to capture the most discriminative global context for WSI classification, we introduce the concept of tokenized global vectors. It is worth mentioning that the importance map for instances can be calculated based on the attention between the tokenized global vector and the embedding of each individual instance. To learn diverse global vectors, we propose two main strategies. First, we push the global vectors toward the centers of the positive bag by a positive instance alignment mechanism. Second, we propose a low-complexity and theoretically guaranteed diversity loss to enforce the orthogonality between the global vectors by utilizing the linear algebra property of the determinantal point process (DPP). In this paper, we explore the design of diverse global representation in the MIL model to model the diversity of instances in WSI. The main contributions are four-fold: (i) We introduce a new perspective on modeling the diversity of instances in WSI. (ii) We further propose a novel MIL aggregation model, termed DGR-MIL, to model diversity in MIL through a set of learnable global vectors. (iii) To learn a diverse global representation (vectors), we propose two main mechanisms: positive instance alignment and a novel diversity loss. 
(iv) Experimental results on two WSI benchmarks demonstrate the proposed DGR-MIL outperforms other competing MIL aggregation methods.

\section{Related Work}
\subsection{Multiple instance learning in WSIs}
MIL has been widely applied in many fields, e.g., pathology~\cite{ilse2018attention,li2021dual,shao2021transmil,zhang2022dtfd}, video analysis~\cite{babenko2010robust,quellec2017multiple}, time series~\cite{early2024inherently,chen2024timemil}.
In particular, the applications of the MIL in Whole Slide Image classification can be roughly summarized into two sub-categories: i) instance-based MIL~\cite{feng2017deep, hou2016patch, xu2019camel} and ii) bag embedding-based MIL. Instance-based methods typically require the propagation of the bag-level label to each of its instances to train the model. Consequently, the final bag-level prediction is obtained by aggregating instance-level predictions. However, empirical studies have proven its performance inferior to the embedding-based competitors because of the noisy instance-level supervision~\cite{wang2018revisiting}. In contrast, bag-embedding-based methods start by projecting instances into feature embeddings and subsequently aggregate the information of these embeddings to obtain the bag-level prediction. Since the introduction of attention-based MIL (AB-MIL)~\cite{ilse2018attention}, the prevailing applications of bag embedding-based MIL in WSI analysis have revolved around this framework. However, AB-MIL operates under the assumption that all instances within a bag are independent and identically distributed while failing to uncover inter-instance correlations. Therefore, numerous of its follow-up works centered around mitigating this limitation by taking advantage of non-local attention mechanism~\cite{li2021dual}, transformer~\cite{shao2021transmil}, pseudo bags~\cite{zhang2022dtfd}, sparse coding~\cite{qiu2023sc}, and low-rank constraints~\cite{lowrankmil}. 

Most existing mainstream MIL methods have modeled correlations mainly through similarity between instances. However, they did not consider the variability of instances between and within bags. Conversely, clustering/prototype-based MIL employs attention scores for selecting prototypes~\cite{scmil,tpmil}, potentially introducing noise and misleading model decisions~\cite{PDL,iterself}. Unlike attention-guided methods, PMIL \cite{pmil} suggests a two-stage framework that first leverages clustering to identify reference prototypes and capture the sub-cluster representation among patch instances and prototypes. However, unrestricted optimization in prototype selection can easily lead to suboptimal outcomes, and a limited number of prototypes can result in a loss of diversity (limited by computational resources).
In this paper, we explicitly model the diversity among instances in bag-embedding-based MIL through a learnable global representation. Although the proposed method falls into the category of transformer-based MILs, it differs from the previous transformer-based MILs~\cite{shao2021transmil, lowrankmil} in two main aspects. First, we model the diversity between instances by comparing instances to the proposed global vectors via a cross-attention mechanism.
Second, we propose a tokenized global vector to summarize the context information of positive instances. 

\subsection{Transformer}
 The transformer~\cite{attentionisallyouneed} has been widely applied in computer vision~\cite{vit,swimtransformer,crossattention2,carion2020end}, time series modeling~\cite{timeserise1,timeserise2}, and the natural language processing fields~\cite{nlp1,nlp2,nlp3}. Standard transformers discover contextually relevant information by modeling the correlation between elements within a sequence through the self-attention mechanism. However, the traditional self-attention operation has quadratic time and space complexity $\mathcal{O}(n^2)$, with respect to a sequence containing $n$ elements. In the context of MIL, sequence length typically becomes quite large since one bag often approximately comprises ten thousand instances. This extremely long sequence poses significant computational intractability. Although \cite{wang2020linformer,shen2021efficient,guo2022beyond} demonstrate that proper approximation of standard self-attention can reduce its quadratic complexity to linear, it still struggles to capture extremely long-term dependencies of context \cite{bhattamishra2020computational,ruoss2023randomized,lowrankmil}.
In contrast, the cross-attention mechanism~\cite{crossattention2,crossattention3}, which was originally proposed to relate positions from one sequence to another, allows models to consider cross-sequence information. Inspired by this, we 
propose to model the diversity between and among instances through a cross-attention between instances and the proposed global vectors (see details in Section~\ref{sec:grmp}). This dramatically reduces the complexity compared to the self-attention mechanism (see Appendix C for details of model complexity) since the number of global vectors is significantly less than the sequence length.

\begin{figure*}[t]
\centering
\includegraphics[width=1.0\textwidth]{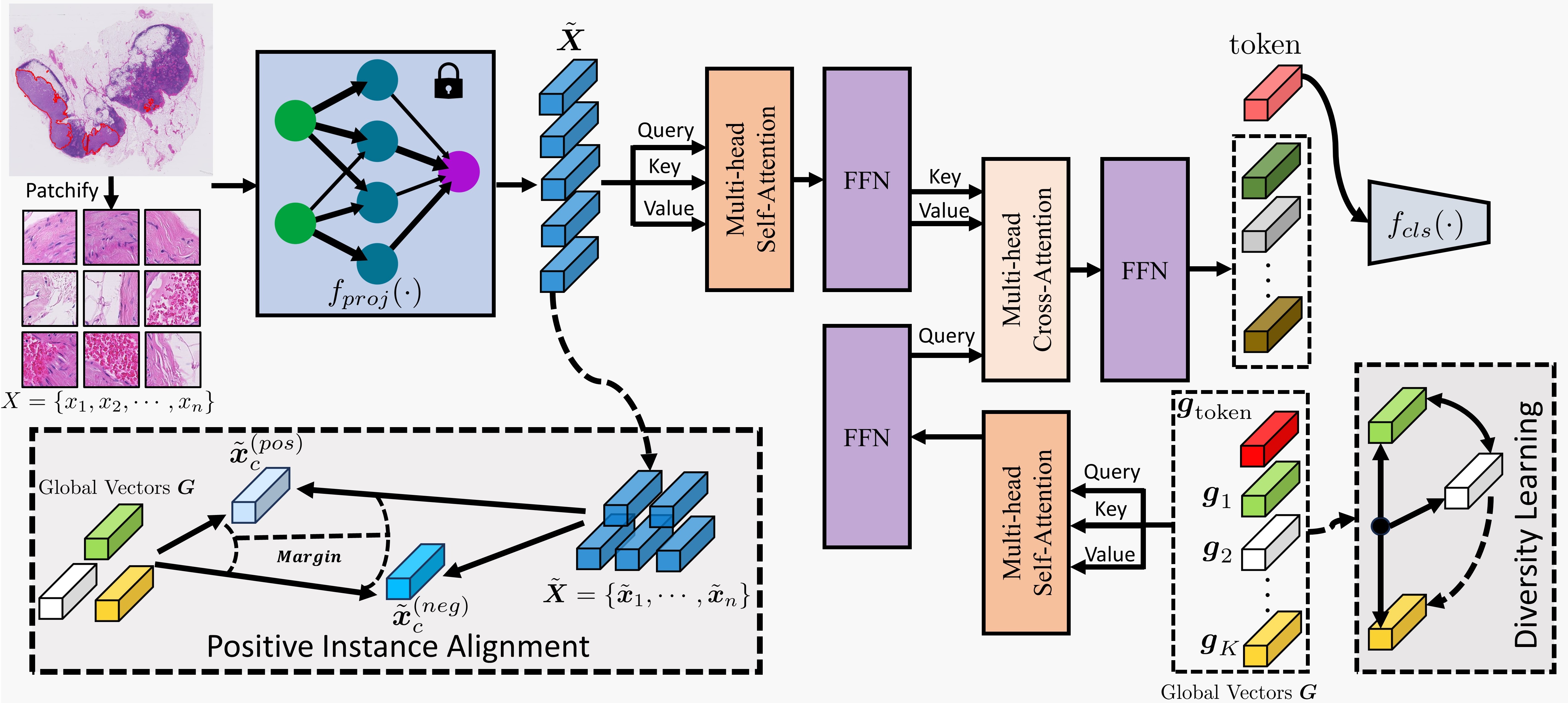} 
\caption{Overview of the proposed DGR-MIL where the global vectors are used for modeling the diversity of instances. The diverse global vectors are learned through the positive instance alignment module and the diversity learning mechanism.}
\label{fig:workflow}
\end{figure*}

\section{Methods}
The proposed DGR-MIL comprises two main parts: i) the design of the global representation in MIL pooling (Section~\ref{sec:grmp}), and ii) the strategy of learning diverse global representation (Section~\ref{sec:GRCL}), where we further propose positive instance alignment and a computational-efficient diversity loss with a theoretical guarantee. The entire framework of DGR-MIL is depicted in Fig. \ref{fig:workflow}.

\noindent \textbf{Preliminary.} Without loss of generality, we take binary MIL classification as an example:
The objective is to predict the bag-level label $Y \in \{0, 1\}$, given a bag of instances $\boldsymbol{X} = \{\boldsymbol{x}_1, \boldsymbol{x}_2, \cdots, \boldsymbol{x}_n\}$, denoting a WSI with $n$ tiled patches. However, the corresponding instance-level labels $\{y_i\}_{i=1}^{n}$ are unknown in most WSI analyses due to the laboriousness of obtaining patch-level annotations. This turns the WSI classification into a weakly-supervised learning scheme according to the standard MIL formulation: 
\begin{equation}
    Y = \begin{cases}
     0, \ \text{iff} \ \sum_{i} y_i = 0 \\
     1, \ \text{otherwise}.
    \end{cases}
\end{equation}
Because of the gigapixel resolution of WSIs, MIL typically cannot be performed in an end-to-end fashion~\cite{clam-sb,end2endsupport,end2endsupport2} and instead necessitates a simplified learning scheme. This simplified MIL learning process comprises three main parts: i) a pre-trained feature extractor $f_{proj}(\cdot)$ that projects each instance into a $L$-dimensional vector, ii) a MIL pooling operator $\sigma(\cdot)$ that combines instance-level embeddings into a bag-level feature, and iii) a bag-level classifier $f_{cls}(\cdot)$ that takes the bag-level feature as input and produces the bag-level prediction as output. Mathematically, this process is given by
\begin{equation}
\begin{split}
     & \ \ \ \ \hat{Y} = f_{cls}(\sigma(\{\Tilde{\boldsymbol{x}}_1, \cdots, \Tilde{\boldsymbol{x}}_n\})), \ \Tilde{\boldsymbol{x}}_i \in \mathbb{R}^{L} \\
     & \text{with} \ \{\Tilde{\boldsymbol{x}}_1, \cdots, \Tilde{\boldsymbol{x}}_n\} = f_{proj}(\{\boldsymbol{x}_1, \cdots, \boldsymbol{x}_n\}),
\end{split}
\end{equation}
where $\hat{Y}$ denotes the predicted bag-level label.
In the attention-based MIL (AB-MIL) \cite{ilse2018attention} framework, the typical formulation for the MIL pooling operator is as follows:
\begin{equation}
    \sigma(\Tilde{\boldsymbol{x}}_i) = \frac{\exp\{\boldsymbol{W}^\mathrm{T} (\text{tanh}(\boldsymbol{V} \Tilde{\boldsymbol{x}}_i)) \odot \text{sigm}(\boldsymbol{U}\Tilde{\boldsymbol{x}}_i) \}}{\sum_{i=1}^{n} \exp\{\boldsymbol{W}^\mathrm{T} (\text{tanh}(\boldsymbol{V} \Tilde{\boldsymbol{x}}_i)) \odot \text{sigm}(\boldsymbol{U}\Tilde{\boldsymbol{x}}_i) \}},
\label{eqn:abmil}
\end{equation}
where $\boldsymbol{W}, \boldsymbol{V}$, and $\boldsymbol{U}$ are learnable parameters. 

\subsection{Global Representation in MIL Pooling}
\label{sec:grmp}

To accommodate the variability of the target lesions within and between bags, we develop a diverse global representation in the MIL pooling stage. Specifically, we define the global representation of the target (positive) instances as a set of learnable vectors given by 
$\boldsymbol{G} = [\boldsymbol{g}_1^\top, \cdots, \boldsymbol{g}_K^\top ]\in \mathbb{R}^{K \times L}$
with $\boldsymbol{g}_k \in \mathbb{R}^{L}$ where $K$ is the number of global vectors. It is worth noting that a feed-forward network (FFN) is used to embed further both the input instance vectors $\Tilde{\boldsymbol{X}}=\{\Tilde{\boldsymbol{x}}_i\}_{i=1}^{n}$ and the global vectors $\boldsymbol{G}$ (see Fig. \ref{fig:workflow}). However, we keep using $\boldsymbol{G}\in \mathbb{R}^{K \times L}$ to denote global vectors for notation brevity.   


\subsubsection{Instance Correlation as Cross Attention.} The standard AB-MIL framework assumes the instances are independent and identically distributed while overlooking the correlation effect between instances. Hence, the self-attention mechanism becomes a natural choice for modeling the inter-instance correlation. However, due to the large number of instances within a bag in MIL, the quadratic time and space complexity $\mathcal{O}(n^2)$ of standard self-attention poses a significant challenge in computation. Alternatively, the previous transformer-based MIL~\cite{shao2021transmil} mitigates this problem by employing Nystrom-Attention~\cite{xiong2021nystromformer}, approximating the standard self-attention with linear complexity, which has proved effective of modeling correlation between positive and negative instances. It could be used to gather similar instances together by attention, benefiting from filtering background information. However, self-attention usage only guarantees the general separation of the positive and negative instances in a bag, which overlooks the diversity between instances and between bags. 

Here, we implicitly model the diversity between instances by comparing the similarity between each instance vector and the proposed diverse global vectors. Specifically, this is achieved through a cross-attention mechanism where the global vector $\boldsymbol{G}$ serves as queries, and a bag of instance vectors $\Tilde{\boldsymbol{X}}$ is used as key-value pairs. Formally, the $h$-th head of the proposed cross attention is given by
\begin{equation}
\begin{split}
    &  \ \ \ \ \ \text{head}_h(\boldsymbol{G}, \Tilde{\boldsymbol{X}}) = \operatorname{Attention}(\boldsymbol{Q}_h, \boldsymbol{K}_h, \boldsymbol{V}_h) \\
    & \boldsymbol{Q}_h = \boldsymbol{G} \boldsymbol{W}_h^{Q}, \ \ \ \boldsymbol{K}_h = \Tilde{\boldsymbol{X}} \boldsymbol{W}_h^{K}, \ \ \ \boldsymbol{V}_h = \Tilde{\boldsymbol{X}} \boldsymbol{W}_h^{V} ,
\end{split}
\end{equation}
where $\boldsymbol{W}^{Q}_h$, $\boldsymbol{W}^{K}_h$, $\boldsymbol{W}^{V}_h \in \mathbb{R}^{L \times L/H}$ are learnable parameters for linear projections, where $H$ is number of heads. For the derivation purposes, we follow the traditional definition of the attention mechanism in the transformer (i.e., $\operatorname{Attention}(\boldsymbol{Q}_h, \boldsymbol{K}_h, \boldsymbol{V}_h) = \operatorname{softmax}\left( \boldsymbol{Q}_h \boldsymbol{K}_h^{\top} / \sqrt{d_k} \right) \boldsymbol{V}_h$). The output of the yielding multi-head cross attention (MHCA) is the concatenation of the outputs from all heads through a linear projection: 
\begin{equation}
    \text{MHCA}(\boldsymbol{G}, \Tilde{\boldsymbol{X}}) = \operatorname{concat}(\text{head}_1; \cdots; \text{head}_H)\boldsymbol{W}^{O},
\label{eqn:mhca}
\end{equation}
where $\boldsymbol{W}^{O} \in \mathbb{R}^{L \times L}$ is a trainable parameter. The proposed cross-attention mechanism reduces the quadratic time and space complexity $\mathcal{O}(n^2)$ in the standard self-attention mechanism to linear $\mathcal{O}(Kn)$ where $K \ll n$. In practice, we applied the Nystrom-Attention to the instance vectors and global vectors before performing the cross-attention (see Fig. \ref{fig:workflow}) for two main reasons. First, applying self-attention to input instance vectors can facilitate filtering out the background. Second, applying self-attention to the global vectors can increase their discrepancies. 

\subsubsection{Tokenized Global Vector.}
The vision transformer includes a class token to encode the globally discriminative representation associated with certain labels in image classification tasks. This token is typically added to the input token embedding by serving as a summary of the entire image. Building upon this inspiration, we propose to add a tokenized global vector $\boldsymbol{g}_{\operatorname{token}}$ as a summary of all the other global vectors. Now, the yielding global vectors can be denoted as  $\Tilde{\boldsymbol{G}}=\{\boldsymbol{g}_{\operatorname{token}}, 
 \boldsymbol{g}_1, \cdots, \boldsymbol{g}_K \} \in \mathbb{R}^{(K+1) \times L}$. The output of the tokenized global vectors after the cross-attention layer (Eq.(\ref{eqn:mhca})) is then used for bag-level classification. Following the convention in AB-MIL, the yielded importance score of each instance can be computed as 
 \begin{equation}
     \sigma(\Tilde{x}_i) = \operatorname{softmax}\left(\frac{(\boldsymbol{g}_{\operatorname{token}} \boldsymbol{W}_{h}^{Q})(\Tilde{\boldsymbol{x}}_i \boldsymbol{W}_{h}^{K})^{\top}}{\sqrt{d_{k}}} \right).
 \end{equation}
 
At first glance, adding the token to the global vectors instead of the input instance embedding appears counterintuitive. However, an in-depth analysis reveals its favorable properties. The proposed global vectors are learned in an unsupervised way (see details in Section~\ref{sec:GRCL}), which poses a significant challenge in perfectly eliminating information from negative instances in the global vectors.  This may be attributed to the similarity between positive instances and their adjacent negative instances, as tumor-adjacent regions typically exhibit high-density, quantitative expression in the spatial relationships of cells~\cite{similaritysupport}. Each diverse global vector encapsulates a collection of analogous tissue features. As a result, certain global vectors emphasize certain types of positive instances. Accordingly, adding tokenized global vectors facilitates the model to capture the most discriminative global representation while suppressing the information from the negative instances (as evident in Fig. \ref{fig:visulization}(b)).

\subsection{Learning Diverse Global Representation}
\label{sec:GRCL}
Due to the weakly-supervised nature of MIL, how to learn the global representation of the target of interest remains an open problem. In this section, we introduce two strategies that can be used to learn a reliable and diverse global representation in MIL, respectively: i) positive instance alignment and ii) diversity learning via utilizing the linear algebra property of the DPP.

\subsubsection{Positive Instance Alignment.} To enforce that the global representation aligns with the instances of interest (i.e., positive instances), we push the global vectors toward the positive bag centers but away from the negative bag centers. To do so, we first define the center of the positive and negative bags as $\Tilde{\boldsymbol{x}}_{c}^{(pos)} \in \mathbb{R}^{L}$ and $\Tilde{\boldsymbol{x}}_{c}^{(neg)} \in \mathbb{R}^{L}$, respectively. Similar to~\cite{he2020momentum}, the positive and negative centers are then updated in a momentum fashion at each training iteration:
\begin{equation}
\begin{split}
    &\Tilde{\boldsymbol{x}}_{c}^{(pos)} = m\Tilde{\boldsymbol{x}}_{c}^{(pos)} + (1- m) \frac{1}{|\mathcal{I}_{pos}|}\sum_{i \in \mathcal{I}_{pos}} \Tilde{x}_i \\
    &\Tilde{\boldsymbol{x}}_{c}^{(neg)} = m\Tilde{\boldsymbol{x}}_{c}^{(neg)} + (1- m) \frac{1}{|\mathcal{I}_{neg}|}\sum_{i \in \mathcal{I}_{neg}} \Tilde{x}_{i}, \\
\end{split}
\end{equation}
where $m$ denotes the momentum update rate, which is set empirically to $0.4$. $\mathcal{I}_{pos}$ and $\mathcal{I}_{neg}$ are the index sets of positive bags and negative bags, respectively. This indicates that the update of the positive instance center occurs only if a positive bag is fed into the network. The same strategy is applied to the negative center update (i.e., updated if and only if a negative bag is encountered). Up to now, we can formulate a set of triplet $\{\boldsymbol{G}, \Tilde{x}_c^{(pos)}, \Tilde{x}_c^{(neg)} \}$. 
The triplet loss~\cite{tripleloss} is then adopted to enforce the global representation $\boldsymbol{G}$ being close to the positive bag center while away from the negative bag center:
\begin{equation}
    \mathcal{L}_{tri} = \sum_{k=1}^{K} [d_+ (G_k, \Tilde{x}_c^{(pos)}) -  d_- (G_k, \Tilde{x}_c^{(neg)}) + \mu]_+,
\end{equation}
where $\mu$ is the margin parameter, and $d$ denotes the distance measure. We use cosine similarity as the distance measure.

\subsubsection{Diversity Learning.}
Although the positive instance alignment mechanism pushes the global representation to be aligned with the positive bag center, it is likely to result in a trivial solution where all the global vectors are identical. However, a diverse global representation is desired to capture the variability of positive instances. Hence, we propose our unique diversity loss inspired by DPP for data selection to maximize the diversity among global vectors and hence better summarize the instances. 
DPP is a well-known diversification tool \cite{kulesza2012determinantal} and is often used to select diverse subsets \cite{chen2018fast,tremblay2019determinantal,derezinski2021determinantal,chen2023rd,10580972}. Inspired so, rather than use it for selection, we utilize it as a diversity measurement.



Mathematically, $\mathcal{P}$ is an L-ensemble DPP if the likelihood of an arbitrary subset $A \subseteq \mathcal{S}$ drawn from the entire set $\mathcal{S}$ satisfies:
\begin{equation}
    \mathcal{P}_{\boldsymbol{L}}(A) \propto \operatorname{det}\left(\boldsymbol{L}_{A}\right),
\label{eqn:dpp}
\end{equation}
where $\boldsymbol{L}_{A}$ denotes a submatrix of the similarity \textit{Gram matrix} $\boldsymbol{L}$ indexed by $A$. In the case of prompting diversity of global vectors $\boldsymbol{G}=[\boldsymbol{g}_1^\top,\cdots,\boldsymbol{g}_K^\top ]$, the similarity matrix is given as $\boldsymbol{L} = \boldsymbol{G} \boldsymbol{G}^{\top} \in \mathbb{R}^{K \times K}$, we simply set $A=\mathcal{B}=[K]$ and each global vector $\boldsymbol{g}_i, ~i \in A$ is treated as a data point, and the total number of subsets can be calculated as $2^{|\mathcal{S}|} = 2^{K}$. It is worth noting that the matrix $\boldsymbol{L}$ is positive semi-definite. 

\begin{lemma} (\cite{kulesza2012determinantal})\label{theorem:geo}
    From a geometric perspective, the determinants in Eq.(\ref{eqn:dpp}) can be interpreted as the squared $|A|$-dimensional volume spanned by its feature vectors:
    \begin{equation}
    \mathcal{P}_{\boldsymbol{L}}(A) \propto \operatorname{det}\left(\boldsymbol{L}_{A}\right) = \operatorname{Vol}^2(\{ \boldsymbol{g}_i \}_{i \in A} ).
\label{eqn:dpp_geo}
\end{equation}

\end{lemma}

\begin{figure}[!t]
    \centering
    \includegraphics[width=0.75\columnwidth]{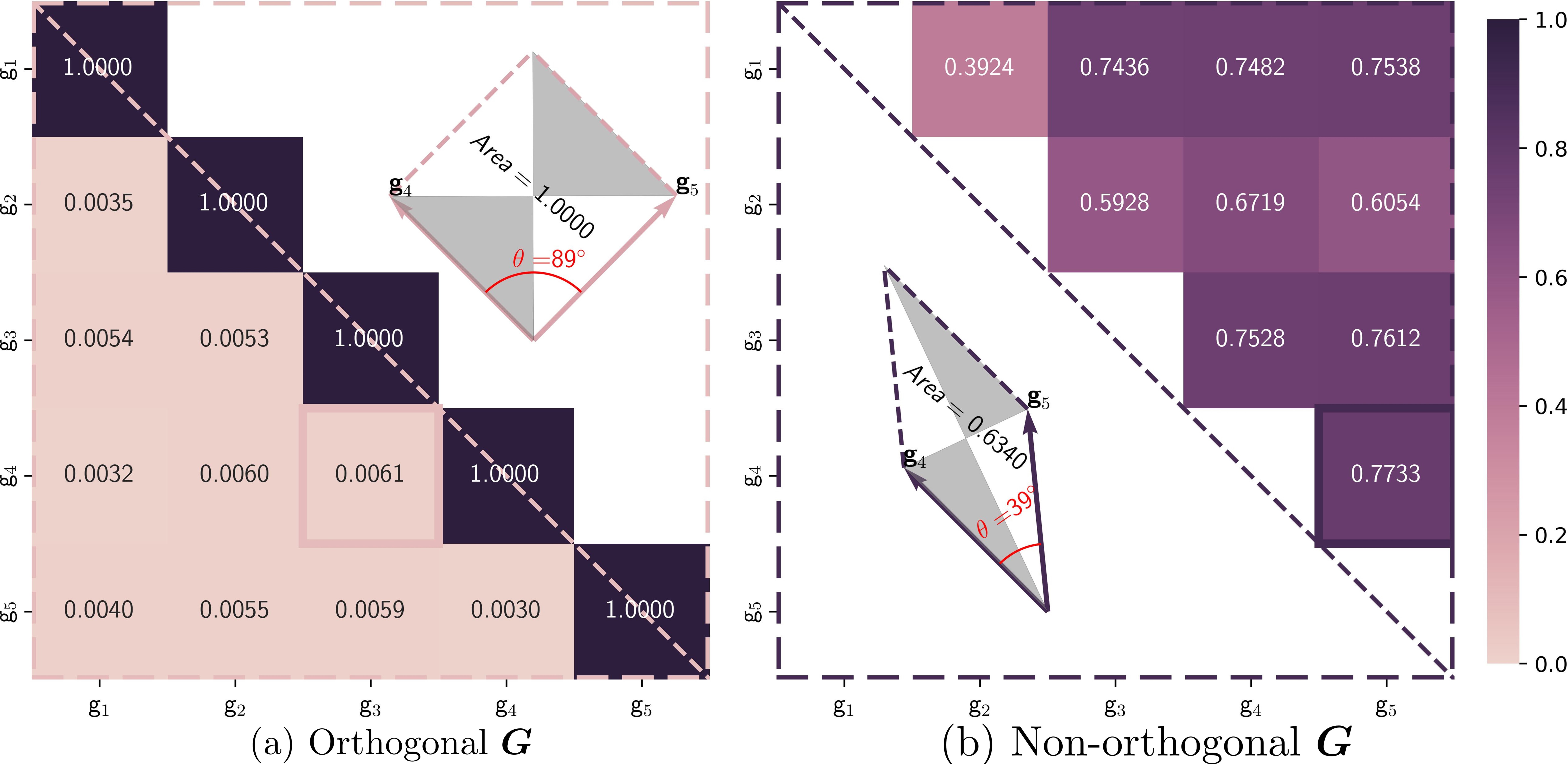}
    \caption{The similarity matrix for the global vectors $\boldsymbol{G}$ learned from the CAMELYON16 dataset in two scenarios: (a) $\boldsymbol{G}$ is orthogonal and (b) $\boldsymbol{G}$ is non-orthogonal. To support Lemma~\ref{theorem:geo} and Remark~\ref{remark:div_loss}, we computed the area of the parallelogram corresponding to the two highly correlated global vectors. We omitted the diagonal elements in subpanel figure (b), as $\boldsymbol{L}_{ii}=1, \ \forall i\in [K]$. }
    \label{fig:diversity}
\end{figure}

Lemma~\ref{theorem:geo} immediately implies that a diverse subset is more likely to span larger volumes. This is because as the similarity between two data points (i.e., $\boldsymbol{L}_{ij: i \neq j}$) increases, they will span fewer areas (see Fig. \ref{fig:diversity}(a) and (b)), hence decreasing the probabilities of sets containing both of them (see Eq.(\ref{eqn:dpp})). Accordingly, feature vectors that are more orthogonal to each other span the largest volumes  (see Fig.\ref{fig:diversity}(a)), hence resulting in the most diverse subsets. 

\begin{theorem}\label{theorem:dpp}
Given a set of global vectors $\boldsymbol{G}=[\boldsymbol{g}_1^\top,\cdots,\boldsymbol{g}_K^\top ]$ with $\|\boldsymbol{g}_i\|=C, \forall i\in [K]$,  maximizing the DPP-based diversity (i.e. $\max \operatorname{det}(\boldsymbol{G}\boldsymbol{G}^{\top})$) results in orthogonal global vectors with $\boldsymbol{g}_i \perp \boldsymbol{g}_j, \  \forall i \neq j,i,j \in [K]$.     
\end{theorem}
\begin{proof}
    The determinant $\operatorname{det}(\boldsymbol{L}) = \operatorname{det}(\boldsymbol{G}\boldsymbol{G}^{\top} )$ is upper-bounded according to Hadamard's inequality~\cite{petersen2008matrix}:
    \begin{equation}
    \begin{split}
         &|\operatorname{det}(\boldsymbol{L})|  \overset{(a)}= \operatorname{det}(\boldsymbol{L}) \overset{(b)} \leq \prod_{i=1}^{K} \boldsymbol{L}_{ii}. \\ 
    \end{split}
    \label{eqn:proof}
    \end{equation}
      \textit{Condition (a)}  is fulfilled because the matrix $\boldsymbol{L}$ is positive semi-definite. The equality of \textit{Condition (b)} is achieved if and only if all non-diagonal entries of $\boldsymbol{G}$ are zeros, meaning rows of the global vectors are orthogonal.
    The normalization constraint in Eq.(\ref{eqn:proof}) leads the upper bound to be the infimum, since  $\boldsymbol{L}_{ii}=\|\boldsymbol{g}\|_i^2\leq C^2$ and it can be achieved if and only if the equality of \textit{Condition (b)} is satisfied.  
    This completes the proof.
\end{proof}
According to Theorem~\ref{theorem:dpp}, we propose a diversity loss $\mathcal{L}_{div}$ to diversify the proposed global vectors by minimizing the negative logarithm of $\operatorname{det}(\boldsymbol{G}\boldsymbol{G}^{\top})$: 
\begin{equation}
    \mathcal{L}_{div}= -\log\operatorname{det}(\boldsymbol{G}\boldsymbol{G}^{\top}),\quad \text{s.t.} \  \|\boldsymbol{g}_i\|=1=C.
    \label{eqn:div_loss}
\end{equation}

\begin{remark}
\label{remark:div_loss}
Theorem~\ref{theorem:dpp} implies that optimal diversity through minimizing our loss is theoretically achievable. This is because enforcing the constraints $\|\boldsymbol{g}_i\|=1$ leads the infimum of $\mathcal{L}_{div}$ to reach zero due to $\log(\boldsymbol{G}\boldsymbol{G}^{\top})_{ii}=\log(\|\boldsymbol{g}_i\|^2)=0$. In contrast, the diversity loss $\mathcal{L}_{div}$ can be arbitrarily small (up to $-\infty$) without the constraint $\|\boldsymbol{g}_i\|=1$, which results in a unstable training.
\end{remark}

We also add a small value $\epsilon=1\times 10^{-10}$ to prevent the logarithm of the determinant from being negative infinity (i.e. any two global vectors become collinear). The final diversity loss is given as 
\begin{equation}
    \mathcal{L}_{div}= -\log\operatorname{det}(\boldsymbol{G}\boldsymbol{G}^{\top}+\epsilon\boldsymbol{I}),
\end{equation}
where $\boldsymbol{I}$ denotes the identity matrix. It is noteworthy that the complexity to compute the loss is approximate $ \mathcal{O}(L)$, which is negligible (see Appendix D).


\subsection{Objective Function}
The proposed MIL model is trained in an end-to-end fashion by jointly optimizing the weighted combination of cross-entropy (ce) loss that corresponds to the bag-level classification, triplet loss, and the proposed diversity loss:
\begin{equation}
    \mathcal{L}_{final} = \mathcal{L}_{ce} + \lambda_{tri} \mathcal{L}_{tri} + \lambda_{div}\mathcal{L}_{div}, 
\end{equation}
where $\lambda_{tri}$ and $\lambda_{tri}$ are balance parameters.


\section{Experiments and Results}
To validate the effectiveness of the proposed DGR-MIL, we conduct experiments on the CAMELYON16 dataset~\cite{bejnordi2017diagnostic} and TCGA-lung cancer dataset (TCGA-NSCLC).

\noindent\textbf{Dataset and Evaluation Metrics.} The two datasets are followed the experimental data partition setting in~\cite{zhang2022dtfd}. For the CAMELYON16, the training set is further divided into training and validation sets with a 9:1 ratio. We report the mean of accuracy, F1 score, and AUC with their corresponding $95\%$ interval on the testing dataset after running five experiments. For the TCGA lung cancer dataset, we perform 4-fold cross-validation experiments, where the dataset is partitioned into training, validation, and testing sets with a patient ratio of 65:10:25.  We report the mean and standard variation of accuracy, F1 score, and AUC on the testing dataset from 4-fold cross-validation.

\noindent\textbf{Experiment Setup.}
Three sets of instance features were extracted using different strategies to evaluate the proposed method's adaptability across various feature embeddings. The first set provided by DTFD-MIL~\cite{zhang2022dtfd}, employing OTSU's method for patch extraction from WSIs and ResNet-50 for feature extraction, resulting in 1024-dimensional vectors per patch. For thorough validation, two additional sets of features were generated by segmenting each WSI into non-overlapping 224x224 patches using threshold filtering, resulting in 3.4 and 10.3 million patches from CAMELYON16 and TCGA lung cancer datasets~\cite{li2021dual,qiu2023sc,lin2023interventional,PDL}, respectively. These patches were processed using ResNet-18 and Vision Transformer, pre-trained on ImageNet, to produce 512 and 768-dimensional feature vectors.

\noindent\textbf{Baseline MIL Models.}
 We compare the proposed model to eight state-of-the-art MIL methods. These models can be roughly divided into two categories: i) AB-MIL~\cite{ilse2018attention} and its variants, including CLAM-SB~\cite{clam-sb}, DS-MIL~\cite{li2021dual}, and DTFD-MIL~\cite{zhang2022dtfd}; ii) the transformer-based methods including Trans-MIL~\cite{shao2021transmil} and ILRA-MIL~\cite{lowrankmil}. iii) clustering/prototype-based MIL including PMIL~\cite{pmil}.

\begin{table*}[t]		
\caption{Main results on the CAMELYON16 dataset and TCGA-NSCLC dataset by using features extracted by different means. Our method statistically outperforms all other competitors (refer to the statistic test in Appendix E)}
\label{tab:experiment_two_benchmark}
		\centering
		\resizebox{1.0\textwidth}{!}{
			\begin{tabular}{lccc|ccc}
				\toprule
                     & \multicolumn{3}{c}{CAMELYON16} & \multicolumn{3}{c}{TCGA-NSCLC} \\
                      \cmidrule(r){2-4} \cmidrule(r){5-7}
				  & Accuracy & F1 &  AUC & Accuracy & F1 &  AUC  \\
				\midrule
				& \multicolumn{6}{c}{\cellcolor{blue!20}\bf ResNet-50 ImageNet Pretrained } \\
                     Classic AB-MIL (\textit{ICML'18})   & $0.845_{(0.839,0.851)}$& $0.780_{(0.769,0.791)}$ & $0.854_{(0.848,0.860)}$ &$0.869_{0.032}$ & $0.866_{0.021}$ & $0.941_{0.028}$\\ %
   
                      DS-MIL (\textit{CVPR'21})  &$0.856_{(0.843,0.869)}$  &$0.815_{(0.797,0.832)}$  &$0.899_{(0.890,0.908)}$  &$0.888_{0.013}$ & $0.876_{0.011}$ &$0.939_{0.019}$\\ %
                    CLAM-SB (\textit{Nature Bio. Eng.'21}) & $0.837_{(0.809,0.865)}$ & $0.775_{(0.755,0.795)}$ & $0.871_{(0.856,0.885)}$ & $0.875_{0.041}$ & $0.864_{0.043}$ & $0.944_{0.023}$\\ %
                    CLAM-MB (\textit{Nature Bio. Eng.'21})  &$0.823_{(0.795,0.850)}$ & $0.774_{(0.752,0.795)}$& $0.878_{(0.861,0.894)} $& $0.878_{0.043} $& $0.874_{0.028}$ & $0.949_{0.019}$\\ %
                    PMIL (\textit{MedIA'23})  &$0.831_{(0.799,0.863)}$& $0.816_{(0.779,0.853)}$ & $0.845_{(0.813,0.876)} $ & $0.873_{0.010}$ &$0.875_{0.011}$ &$0.933_{0.007}$ 
                    \\%
                    Trans-MIL (\textit{NeurIPS'21}) & $0.858_{(0.848,0.868)}$  & $0.797_{(0.776,0.818)}$  & $0.906_{(0.875,0.937)}$ & $0.883_{0.022}$  & $0.876_{0.021}$  & $0.949_{0.013}$ \\ %
                    
                    DTFD-MIL (MaxS) (\textit{CVPR'22})  &$0.864_{(0.848,0.880)} $& $0.814_{(0.802,0.826)}$& $0.907_{(0.894,0.919)} $&$ 0.868_{0.040 }$& $0.863_{0.029}$& $0.919_{0.037}$
                    \\ %
                    DTFD-MIL (MaxMinS) (\textit{CVPR'22})  &$0.899_{(0.887,0.912)}$ & $0.865_{(0.848,0.882)}$ & $0.941_{(0.936,0.944)}$ & $0.894_{0.033}$ & $0.891_{0.027}$ & $0.961_{0.021}$ \\ %
                    DTFD-MIL (AFS) (\textit{CVPR'22})  & $0.908_{(0.892,0.925)}$ & $0.882_{(0.861,0.903)}$ & $0.946_{(0.941,0.951)}$ & $0.891_{0.033}$ & $0.883_{0.025}$ & $0.951_{0.022}$
                    \\%
                    ILRA-MIL (\textit{ICLR'23}) 
                    &$0.848_{(0.844,0.853)}$& $0.826_{(0.823,0.829)}$ & $0.868_{(0.852,0.883)} $  & $0.895_{0.017}$ &$0.896_{0.017}$ &$0.946_{0.014}$  
                      \\                 
                     \rowcolor{blue!8} 
                     \textbf{Our}  
&$\mathbf{0.917}_{(0.902, 0.931)}$& $\mathbf{0.913}_{(0.898, 0.928)}$ & $\mathbf{0.957}_{(0.951, 0.963)}$ & $\mathbf{0.908}_{0.015}$&$\mathbf{0.911}_{0.018}$&$\mathbf{0.963}_{0.008}$

\\ 
                     
                 \hline
                    & \multicolumn{6}{c}{\cellcolor{blue!20}\bf ResNet-18 ImageNet Pretrained} \\
                    Classic AB-MIL (\textit{ICML'18})  & $0.805_{(0.772,0.837)}$& $0.786_{(0.757,0.815)}$ & $0.843_{(0.827,0.858)}$ &$0.874_{0.005}$ & $0.873_{0.006}$ & $0.937_{0.001}$\\ %
   
                    DS-MIL (\textit{CVPR'21})  &$0.791_{(0.739,0.843)}$  &$0.776_{(0.712,0.840)}$  &$0.814_{(0.754,0.875)}$  &$0.831_{0.012}$ & $0.838_{0.008}$ &$0.896_{0.009}$\\ %
                    CLAM-SB (\textit{Nature Bio. Eng.'21})  & $0.792_{(0.769,0.815)}$ & $0.766_{(0.746,0.786)}$ & $0.811_{(0.777,0.845)}$ & $0.869_{0.010}$ & $0.869_{0.010}$ & $0.931_{0.006}$\\ %
                    CLAM-MB (\textit{Nature Bio. Eng.'21}) &$0.786_{(0.754,0.818)}$ & $0.770_{(0.746,0.795)}$& $0.825_{(0.808,0.843)} $& $0.880_{0.016} $& $0.880_{0.016}$ & $0.944_{0.012}$\\ %
                    PMIL (\textit{MedIA'23})  &$0.800_{(0.775,0.825)}$& $0.784_{(0.765,0.804)}$ & $0.829_{(0.807,0.851)} $ 
                     & $0.856_{0.006}$ &$0.862_{0.003}$ &$0.933_{0.010}$\\%
                    Trans-MIL (\textit{NeurIPS'21}) & $0.839_{(0.822,0.856)}$  & $0.827_{(0.805,0.848)}$  & $0.854_{(0.823,0.886)}$ & $0.877_{0.009}$  & $0.879_{0.008}$  & $0.938_{0.014}$ \\ %
                    
                    DTFD-MIL (MaxS) (\textit{CVPR'22})  &$0.856_{(0.824,0.887)} $& $0.792_{(0.742,0.842)}$& $0.878_{(0.862,0.893)} $&$ 0.830_{0.014 }$& $0.821_{0.020}$& $0.893_{0.015}$
                    \\ %
                    DTFD-MIL (MaxMinS) (\textit{CVPR'22})  &$0.833_{(0.807,0.858)}$ & $0.768_{(0.747,0.788)}$ & $0.878_{(0.872,0.883)}$ & $0.853_{0.012}$ & $0.850_{0.021}$ & $0.925_{0.013}$ \\ %
                    DTFD-MIL (AFS) (\textit{CVPR'22})  & $0.817_{(0.791,0.843)}$ & $0.734_{(0.687,0.781)}$ & $0.868_{(0.841,0.896)}$ & $0.870_{0.007}$ & $0.864_{0.012}$ & $0.935_{0.010}$
                    \\%
                    ILRA-MIL (\textit{ICLR'23})  &$0.831_{(0.768,0.895)}$& $0.819_{(0.768,0.871)}$ & $0.852_{(0.811,0.893)} $ & $0.878_{0.002}$ &$0.879_{0.001}$ &$0.937_{0.004}$  

\\                  
                     \rowcolor{blue!8}
                     \textbf{Our} 
&$\mathbf{0.873}_{(0.862, 0.884)}$& $\mathbf{0.862}_{(0.852, 0.871)}$ & $\mathbf{0.898}_{(0.886, 0.909)}$ & $\mathbf{0.891}_{0.029}$&$\mathbf{0.890}_{0.021}$&$\mathbf{0.955}_{0.023}$

\\ 
                    
                \hline
                 & \multicolumn{6}{c}{\cellcolor{blue!20}\bf Vision Transformer ImageNet Pretrained } \\
                     Classic AB-MIL (\textit{ICML'18})   & $0.851_{(0.837,0.865)}$& $0.835_{(0.810,0.860)}$ & $0.873_{(0.840,0.906)}$ &$0.904_{0.011}$ & $0.904_{0.010}$ & $0.953_{0.013}$\\ %
   
                    DS-MIL (\textit{CVPR'21})  &$0.810_{(0.741,0.879)}$  &$0.806_{(0.742,0.869)}$  &$0.871_{(0.836,0.906)}$  &$0.875_{0.020}$ & $0.879_{0.016}$ &$0.933_{0.016}$\\ %
                    CLAM-SB (\textit{Nature Bio. Eng.'21}) & $0.839_{(0.831,0.847)}$ & $0.816_{(0.799,0.834)}$ & $0.864_{(0.841,0.887)}$ & $0.907_{0.008}$ & $0.907_{0.001}$ & $0.954_{0.014}$\\ %
                    CLAM-MB (\textit{Nature Bio. Eng.'21}) &$0.826_{(0.806,0.846)}$ & $0.804_{(0.795,0.813)}$& $0.851_{(0.825,0.878)} $& $0.911_{0.007} $& $0.911_{0.007}$ & $0.959_{0.008}$\\ %
                    PMIL (\textit{MedIA'23})  &$0.843_{(0.831,0.856)}$& $0.826_{(0.814,0.838)}$ & $0.843_{(0.820,0.867)} $ 
                     & $0.882_{0.009}$ &$0.884_{0.006}$ &$0.940_{0.006}$
                     \\
                    Trans-MIL (\textit{NeurIPS'21}) & $0.862_{(0.841,0.883)}$  & $0.846_{(0.823,0.869)}$  & $0.860_{(0.848,0.873)}$ & $0.909_{0.009}$  & $0.909_{0.009}$  & $0.953_{0.006}$ \\ %
                    
                    DTFD-MIL (MaxS) (\textit{CVPR'22}) & $0.846_{(0.832,0.860)}$& $0.767_{(0.746,0.787)}$ & $0.859_{(0.842,0.876)}$ &$0.904_{0.011}$ & $0.904_{0.010}$ & $0.953_{0.013}$
                    \\
                    DTFD-MIL (MaxMinS) (\textit{CVPR'22})  &$0.839_{(0.826,0.851)}$ & $0.752_{(0.742,0.763)}$ & $0.862_{(0.836,0.888)}$ & $0.895_{0.013}$ & $0.892_{0.016}$ & $0.952_{0.011}$ \\ %
                    DTFD-MIL (AFS) (\textit{CVPR'22})  & $0.831_{(0.818,0.844)}$ & $0.759_{(0.737,0.781)}$ & $0.880_{(0.864,0.897)}$ & $0.901_{0.005}$ & $0.900_{0.008}$ & $0.959_{0.012}$
                    \\%
                    ILRA-MIL (\textit{ICLR'23})  &$0.850
_{(0.825
,0.875)}$& $0.838
_{(0.812
,0.865
)}$ & $0.864
_{(0.843
,0.885
)} $ & $0.902_{0.007}$ &$0.904_{0.007}$ &$0.954_{0.006}$

                     \\
                     %
                     \rowcolor{blue!8}
                     \textbf{Our}  
                    & $\mathbf{0.893}_{(0.889,0.897)}$ & $\mathbf{0.882}_{(0.877,0.886)}$ & $\mathbf{0.891}_{(0.884,0.899)}$ & $\mathbf{0.926}_{0.008}$ & $\mathbf{0.925}_{0.008}$ & $\mathbf{0.969}_{0.004}$

                     \\ 
                     
				\bottomrule
 			\end{tabular}}

   \vspace{-0.5cm}
\end{table*}
\noindent\textbf{Implementation Details.}  All the models are trained using the parameter settings provided by~\cite{shao2021transmil,zhang2022dtfd,li2021dual,clam-sb,lowrankmil}. (See Appendix B, including our method).

\noindent\textbf{Additional Experiments.} We also include the experiments on using CTransPath \cite{wang2022transformer} as feature extractor for CAMELYON16 dataset. Additionally, to validate the generalizability of our method on broader applications other than WSI, we conduct the experiment on MIL benchmark \cite{dietterich1997solving,andrews2002support}. Our method demonstrates the obvious superiority over other methods in both experiments. Please refer to Appendix F.


\subsection{Experimental Results}
The proposed method outperforms the other state-of-the-art MIL aggregation models by a large margin in both the CAMELYON16 and TCGA-NSCLC datasets using features extracted by three different means (see Table~\ref{tab:experiment_two_benchmark}). We also show the statistical superiority of our method in Appendix E. Specifically, the proposed model outperforms the second-best models in terms of accuracy (1.7\%; 1.3\%), F1 score (3.1\%; 1.5\%), and AUC (1.1\%; 1.7\%) when using features extracted from ResNet-50 in CAMELYON16 and TCGA-NSCLC, respectively. 
A similar performance gain is observed on features extracted from ResNet-18 including accuracy (3.4\%; 1.1\%), F1 score (3.5\%; 1.0\%), and AUC (4.4\%; 1.1\%). We also observe an improvement in accuracy (3.4\%; 1.1\%), F1 score (3.5\%; 1.0\%), and AUC (4.4\%; 1.1\%) when using features extracted from the vision transformer. In general, the proposed model shows a greater performance improvement in the CAMELYON16 dataset compared to the TCGA-NSCLC dataset. This might be attributed to the fact that CAMELYON16 consists of more diverse instances than TCGA-NSCLC. 

We also observe the performance of the three sets of feature embeddings varied: the ViT feature embeddings outperform the ResNet-18 features but show inferior performance compared to the ResNet-50 features. This is mainly attributed to the fact that a greater number of positive instances is extracted by the ResNet-50 (provided by DTFD-MIL) as shown in Fig. \ref{fig:Ablation}(d). In contrast, a smaller portion of positive instances in the extracted patches may accompany a drop in performances~\cite{plp2}. This phenomenon benefits the pseudo-bag partitions in DTFD-MIL, as more positive instances within a bag are prone to result in less noisy pseudo-bag labels. This accounts for the drop in DTFD-MIL performance when applied to feature embeddings that contain a lower proportion of positive instances.

\subsection{Ablation Studies}
We conduct ablation studies on model design variants in the CAMELYON16 dataset with features extracted by a ResNet-50, unless specified otherwise. 
\begin{figure*}[!t]
\centering
\includegraphics[width=1.0\textwidth]{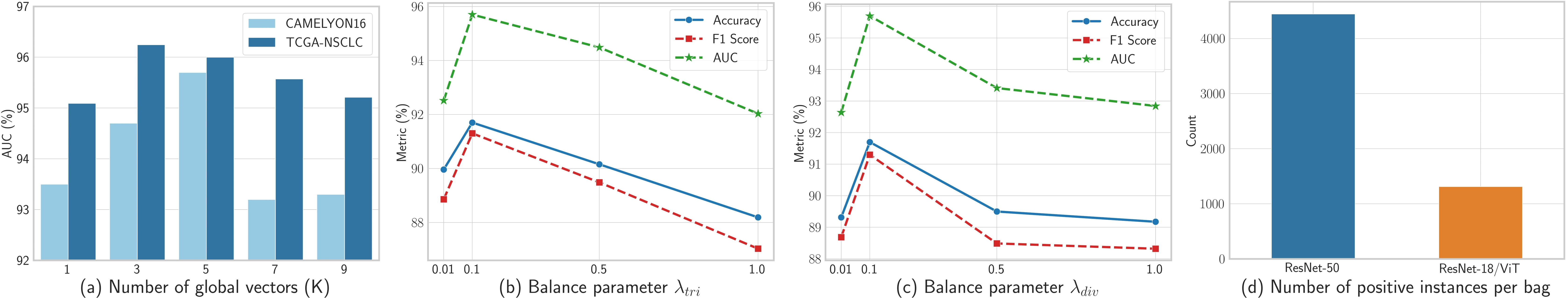} 
\caption{Ablation studies on (a) number of non-tokenized global vectors on both CAMELYON16 and TCGA-NSCLC datasets, (b) and (c) balance parameter $\lambda_{tri}$ and $\lambda_{div}$ on CAMELYON16 dataset, respectively. (d) Comparison in the number of positive instances per bag.}
\label{fig:Ablation}
\end{figure*}




\begin{table}[!t]
\vspace{-0.2cm}
    \centering
    \begin{minipage}{0.45\columnwidth}
        
        \centering
\caption{The ablation studies on different modules. $\mathcal{P}$: Positive instance alignment module. $\mathcal{D}$: Diversity loss.}
\label{tab:aba_table1}
\resizebox{1\textwidth}{!}{%
\begin{tabular}{ll|ccc|ccc}
\toprule %
 $\mathcal{P}$ & $\mathcal{D}$ & \multicolumn{3}{c}{CAMELYON16} & \multicolumn{3}{c}{TCGA-NSCLC} \\
                      \cmidrule(r){3-5} \cmidrule(r){6-8}
				&  & Accuracy & F1 &  AUC & Accuracy & F1 & AUC  \\
				\midrule
\ding{55} & \ding{55} & \textcolor{gray}{0.895}  & \textcolor{gray}{0.887} & \textcolor{gray}{0.922}  & \textcolor{gray}{0.872} & \textcolor{gray}{0.875} & \textcolor{gray}{0.928}\\ 
\ding{55} &\ding{51}  & 0.906 & 0.900 & 0.938 & 0.896 & 0.896 & 0.952 \\
  \ding{51}& \ding{55} & 0.917 & 0.910 & 0.944 & 0.900 & 0.904 & 0.956 \\
 \ding{51}& \ding{51} & 0.917 & 0.913 & 0.957 & 0.908 & 0.911 & 0.963 \\

\bottomrule
\end{tabular}%
}
    \end{minipage}%
    \hfill
    \begin{minipage}{0.5\columnwidth}
    \centering
    \caption{ The ablation studies on tokenized global representation.}
\label{tab:token}
\resizebox{1\textwidth}{!}{%
\begin{tabular}{c|ccc|ccc}
\toprule %
 $\boldsymbol{g}_{\operatorname{token}}$ & \multicolumn{3}{c}{CAMELYON16} & \multicolumn{3}{c}{TCGA-NSCLC} \\
                      \cmidrule(r){2-4} \cmidrule(r){5-7}
				  & Accuracy & F1 &  AUC & Accuracy & F1 &  AUC  \\
				\midrule

 \ding{55} & 0.907 & 0.900 & 0.935 & 0.903 & 0.905 & 0.957 \\
 \ding{51} & 0.917 & 0.913 & 0.957 & 0.908 & 0.911 & 0.963 \\ 
 \bottomrule
\end{tabular}%
}
\vspace{-0.2cm}
    \end{minipage}
\end{table}

\noindent\textbf{Effectiveness of the Proposed Global Representation.} 
We ablate different components of the proposed model, i.e., the positive instance alignment module and the diversity loss. While the model without these two components serves as the baseline in Table~\ref{tab:aba_table1}.
We first observe that incorporating the proposed global vectors described in Section~\ref{sec:grmp}
(without employing any of the learning strategies in Section~\ref{sec:GRCL}) yielded an AUC of 0.922 and 0.928. This AUC exceeds that of most existing MIL models, except for DTFD-MIL (MaxMinS \& AFS) (see Table~\ref{tab:experiment_two_benchmark} and~\ref{tab:aba_table1}). 
Subsequently, by including the proposed positive instance alignment module, we observe a performance gain of (2.2\%, 2.8\%) in accuracy, (2.3\%, 2.9\%) in F1 score, and (2.2\%, 2.8\%) in AUC. Up to now, we outperform the DTFD-MIL in terms of accuracy and F1 score (see Table~\ref{tab:experiment_two_benchmark} and~\ref{tab:aba_table1}), and achieve a similar AUC (AUC = 0.944,0.956) compare to the DTFD-MIL(AFS) (AUC = 0.946,0.951). Further incorporating the proposed diversity loss into the objective function yields a performance gain of (1.3\%,0.7\%) in AUC, which outperforms DTFD-MIL (AFS) by (1.1\%,1.2\%).

\begin{figure*}[!t]
\centering
\includegraphics[width=0.95\textwidth]{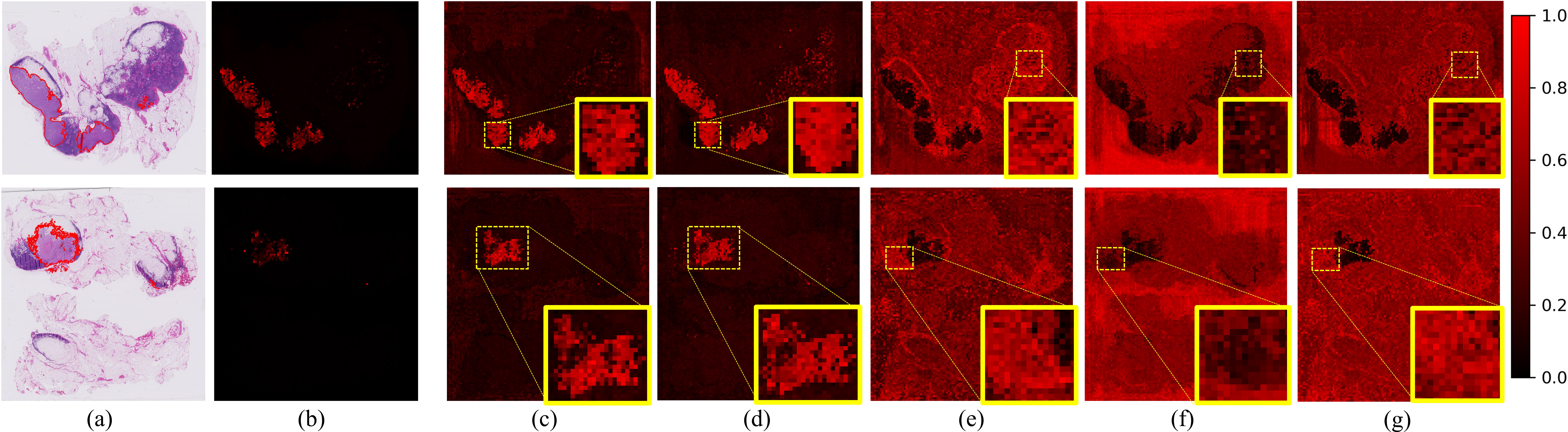} 
\caption{Visualization of the attention map: (a) raw WSI with the ground-truth annotation, (b) the attention map computes using the tokenized global vectors, and (c-g) the attention map computes using the other ($K-1$) global vectors with $K=6$ in our experiment. }
\label{fig:visulization}
\vspace{-0.3cm}
\end{figure*}

\noindent\textbf{Effectiveness of the Tokenized Global Representation.} As shown in Table \ref{tab:token}, including the tokenized global vector $\boldsymbol{g}_{\operatorname{token}}$ yields a remarkable performance gain by improving accuracy by (1.0\%, 0.5\%), F1 score by (1.3\%, 0.6\%), and AUC by (2.2\%, 0.6\%). As consistent with the pathological findings that instances are diverse, we observe that different global vectors indeed corresponded to different instance representations, which can be depicted by the attention map produced by different global vectors in Fig. \ref{fig:visulization}. 
However, we also observe that the learned global vectors still include non-tumor related representation, particularly around tumor boundaries, as positive instances around tumor boundaries have a similar appearance to surrounding negative instances (see Fig. \ref{fig:visulization}.(c) and (d)). As a result, incorporating tokenized global vectors can mitigate this problem by capturing the most discriminative positive (tumor) regions (see Fig. \ref{fig:visulization}.(b)).

\noindent\textbf{Number of Global Vectors.} We find that the optimal number of global vectors $K$ in different data sets may vary due to dataset intrinsic properties. 
Specifically, the optimal $K$ for the CAMELYON16 and TCGA-NSCLC dataset are $K=5$ and $K=3$, respectively (Fig. \ref{fig:Ablation}.(a)). 
We observe that an overly large $K$ is likely to decrease performance as it will harden the learning task (see Fig. \ref{fig:Ablation}.(a)). 

\noindent\textbf{Loss Balance Hyperparameters.} 
By conducting a grid search, we find that the optimal setting of the balance parameters is $\lambda_{tri} = 0.1$ and $\lambda_{div} = 0.1$ (see Fig. \ref{fig:Ablation}.(b) and (c)). An overly small $\mathcal{L}_{tri}$ and $\mathcal{L}_{div}$  (e.g., $0.01$) is likely to enforce inadequate constraints on the learned global representation by deviating it from learning meaningful information of instance of interest. While larger balance parameters (e.g., $\{0.5, 1.0\}$) distract the model from the main classification task, leading to a drop in classification performance.

\section{Conclusion}
Inspired by the pathological fact that instances are diverse, we propose a novel MIL model from the perspective of modeling diversity in instances through the cross-attention between instances and a set of learnable and diverse global vectors. To learn the global vectors, we propose a positive instance alignment mechanism and the DPP-driven diversity loss. Extensive experiments demonstrate that the proposed MIL model competed favorably against other existing MIL models. Importantly, our work provides an explicit way to account for the diversity in WSI. This pathology-driven approach is beneficial in capturing heterogeneity among the patient population. We also narrowed the performance gap between the diversity-drive MIL method and mainstream MIL.

\section{Acknowledgement} This work was partially supported by the grants from NIH (R01EY032125, and R01DE030286), and the State of Arizona via the Arizona Alzheimer Consortium.

\bibliographystyle{splncs04}
\bibliography{ref}

\clearpage 
\appendix

 \clearpage\section*{\Large Supplementary Materials - DGR-MIL: Exploring Diverse Global Representation in Multiple Instance Learning for Whole Slide Image Classification}

\input{supplementary}


\end{document}

%% file: supplementary.tex
\section{Measuring Diversity Based on Rate-distortion Theory}
\begin{figure*}[!ht]
    \centering
    \includegraphics[width=0.88\columnwidth]{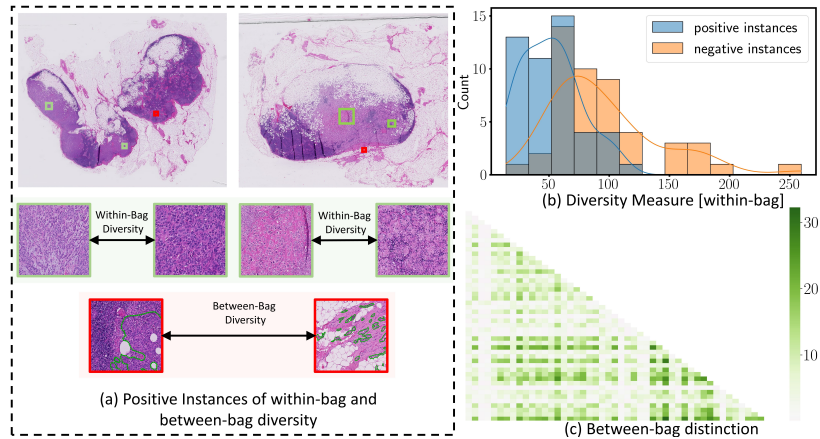}
    \caption{(a) Examples of positive instances of with-bag and between-bag diversities measured by rate-distortion theory. (b) Histogram of the diversity measure within positive bags on the CAMELYON16 dataset. (c) The between-bag distinction measures the pair-wise similarity between bags.}  
    \label{fig:1}
\end{figure*}
 Rate-distortion (RD) theory is a fundamental concept in \textit{information theory} to describe the lossy compression for arbitrary data sources with tolerable distortion. Here, rate $R$ refers to the number of bits or units per symbol of information required to represent the source data or signal; while distortion measures the quality of the reconstructed data compared to the original source data. 
 Mathematically, given an arbitrary source $X$, we can use finite bits $nR$ bits to encode a sequence of $n$ samples $X^n$ with $f_n(X^n)$ using a size codebook $2^{nR}$, and then decode it with
$\hat{X^n}=g_n(f_n(X^n))$. Accordingly, the reconstruction error for the sample sequence $x^n$ can be computed as $d(x^n,\hat{x}^n) := 1/n \sum_{i=1}^n d(x_i,\hat{x}_i)$ for some distance measure $d(\cdot)$. The most commonly used distortion metric is Mean Squared Errors (MSE), which is presented as $\epsilon^2:=1/n \sum_{i=1}^n (x_i-\hat{x}_i)^2$ and distortion $D$ is defined as $D:=\mathbb{E}[d(X^n, \hat{X}^n)]$ \cite{cover1999elements}. The rate $R$ is computed for a sequence with infinite length ($n \rightarrow \infty$) and distortion $D$. For a Gaussian source, given a finite dataset $\boldsymbol{X}=[\boldsymbol{x}_1,\boldsymbol{x}_2,\cdots,\boldsymbol{x}_n]\in \mathbb{R}^{d\times n}$, the theoretical coding rate with a small tolerable MSE distortion $\epsilon^2$, can be approximately estimated
as \cite{ma2007segmentation}, 
\begin{align}\label{eq:r0}
    R(\boldsymbol{X}, \epsilon) &:= \frac{1}{2} \log \operatorname{det}\left(\boldsymbol{I}+\frac{d}{n \epsilon^2} \boldsymbol{X} \boldsymbol{X}^{\top}\right),
\end{align}
where the unit of $ R(\boldsymbol{X}, \epsilon)$ is bit/dimension or nat/dimension for log base $2$ or $e$, respectively. Accordingly, the rate of the sub-space for each class $i$ can be approximated,
\begin{align}\label{eq:subclass}
       R^c_i(\boldsymbol{X}, \epsilon \mid{C}_i) :=\frac{1}{2} \log \operatorname{det}\left(\boldsymbol{I}+\frac{d}{|C_i| \epsilon^2} \boldsymbol{X}_{C_i} \boldsymbol{X}^{\top}_{C_i}\right) ,
\end{align}
where $C_i$ is the index set of class $i$,
$c_T$ is the number of classes, $\boldsymbol{X}_{C_i}$ is a matrix using columns of $\boldsymbol{X}$ indexed by $C_i$ ($\boldsymbol{X}[:,C_i]$), and $|C_i|$ is the cardinality of $C_i$. 
Having adopted the assumption in \cite{yu2020learning}, we use the latent features extracted by the projector to estimate the diversity.

To better illustrate the way to compute the diversity, we copy Fig.\ref{fig:1} from the main body to here. In Fig.\ref{fig:1}.(b), we use Eq. \ref{eq:subclass} to compute the within-bag diversity, which refers to either all negative instances or positive instances (if applicable) from the same bag. The instances are treated as a data matrix $ \boldsymbol{X}$ in Eq.\ref{eq:r0}. We separately compute the diversity for each bag from the test set (80 negative bags and 49 positive bags), which results in a total of 129 negative within-bag diversity data points and 49 positive within-diversity data points. Then, we plot the histograms in Fig.\ref{fig:1}.(b), where the x-axis denotes the measure of diversity and the y-axis denotes the count (or frequency) of the diversity within the interval (i.g. the width and height in a bin, respectively).
It evidences both positive and negative within-bag instances are diverse and on a comparable scale. In Fig.\ref{fig:1}.(c), we use the rate reduction from \cite{yu2020learning} to compute the between-bag distinction of positive instances for every two bags. A rate reduction is presented as
\begin{align}
    \Delta R:=R(\boldsymbol{X}[:,C_1\cup C_2], \epsilon)- \sum_{i=1}^{2}\frac{|C_i|}{n} {R^c_i(\boldsymbol{X}, \epsilon \mid {C}_i)},
\end{align}
where $C_1$ and $ C_2$ are the index sets of two sub-space. This concept is used to describe the difference to encode the entire space and encode the sum of all sub-spaces, and a higher value indicates two sub-spaces are more discriminative; hence, we employed it as a metric to describe the distinction between two bags. In detail, we compare the distinction for every two positive bags. In each computation, 
$C_1$ and $ C_2$ denote the indices of positive instances from two different bags, respectively. $C_1\cup C_2$ denotes all instances from the selected two bags. Fig.\ref{fig:1}.(c) denotes the pair-wise distinction matrix, and we neglected the diagonal elements since they are zero. We also neglected the upper-triangle elements since this distinction matrix is symmetric.  

\begin{table*}[t]
    \centering
    \caption{All training parameters setting for all methods in experiments. Here, Cosine annealing* denotes cosine decay with 20 epoch linear warmup from 1e-5. AMP represents automatic mixed precision, and the grad clip was clipped gradient norm constrained of model weight. Here, BCE was BCEWithLogitsLoss, which combines a sigmoid layer and the binary cross entropy loss.}
    \resizebox{0.99\textwidth}{!}{
    \begin{tabular}{l|l|l|l|l|l|l|l}\toprule
        \multicolumn{1}{c|}{\textbf{Parameters Setting}} & \textbf{AB-MIL} & \textbf{CLAM-SB/MB}& \textbf{DS-MIL}& \textbf{DTFD-MIL} & \textbf{Trans-MIL}  & \textbf{ILRA-MIL}  & \textbf{Our proposed Method }\\ \hline
        optimizer & Adam    &    Adam  &   Adam  &   Adam  &  Radam  & Adam & SGD   \\\cline{1-8}
        learning rate  &  1e-3    & 1e-4 &    1e-4  &   1e-4  &   2e-4  & 1e-4 & 5e-4   \\\cline{1-8}
        weight decay  &    0.005   & 1e-5 &   5e-3 &   1e-4 & 1e-5 & 1e-4 & 1e-4\\   \cline{1-8}                
        scheduler  &   Cosine  annealing* &  Cosine  annealing*& Cosine annealing &  MultiStepLR &   LookAhead~\cite{lookahead}  & Cosine annealing & Cosine annealing*\\ \cline{1-8}
        Dropout rate &           0.15    &           0.15 &   0.15  &   0.15 & 0.15  & 0.15  & 0.15    \\\cline{1-8}
        epoch  &  200  & 200  &   200 &   200    &   200  &   200 & 200 \\ \cline{1-8} 
        loss  &  BCE &  BCE &   BCE   &   BCE +  Tier-2 loss &  BCE   &  BCE   & $ \mathcal{L}_{ce} + \lambda_{tri} \mathcal{L}_{tri} + \lambda_{div}\mathcal{L}_{div}$\\ \cline{1-8}
        other settings  &  None  &  Early stop  &   Droppath = 0.2 &   grad clip = 5 & AMP & Xavier initialize & Warmup training strategy  \\ \cline{1-8} 
    \end{tabular}
    }
    \label{parameter}
    
\end{table*}

\section{Baseline Models Parameter Setting}
The baseline MIL methods include AB-MIL~\cite{ilse2018attention}, CLAM-SB, multi-attention CLAM-MB~\cite{clam-sb}, DS-MIL~\cite{li2021dual}, DTFD-MIL~\cite{zhang2022dtfd}, Trans-MIL~\cite{shao2021transmil} and ILRA-MIL~\cite{lowrankmil}. We follow the optimal parameter settings outlined in their original papers. The detailed parameters that we use to train all the baselines and the proposed model are shown in Table~\ref{parameter}. It is worth noting that our method adopts the linear learning rate warmup for the first 20 epochs, and details can be referred to~\ref{warmuptrainingstrategy}.

\subsection{ResNet-50 ImageNet Pre-Trained}
We use extracted features released by the DTFD-MIL. Each patch was embedded into a 1024-dimensional vector using a ResNet-18 pretrained on ImageNet~\cite{resnet}. The instance features are directly fed to MIL methods for training. In the experiments, we consistently set the middle layer (Some MIL methods including feed-forward layers before entering the aggregation method) output dimension to 512, For example, TransMIL~\cite{shao2021transmil},  ILRA~\cite{lowrankmil}, DTFD~\cite{zhang2022dtfd}, ABMIL~\cite{ilse2018attention}, and the proposed method.

\subsection{ResNet-18 ImageNet Pre-Trained}
\label{resnet18}
Different from DTFD-MIL, we employ the threshold filter method (entropy $<$ 5 discarded) to extract patches from raw WSIs~\cite{li2021dual}.  This results in fewer patches compared to DTFD-MIL. Each patch was embedded into a 512-dimensional vector as an instance feature. Here, we consistently set the middle layer output dimension to 256 in all MIL methods, including ILRA~\cite{lowrankmil}, DTFD~\cite{zhang2022dtfd}, ABMIL~\cite{ilse2018attention}, and the proposed method. Here, the TransMIL middle layers dimension output is 512, following the settings in its original paper~\cite{shao2021transmil}. Here, The TransMIL middle layers dimension output was 512, following the original paper setting~\cite{shao2021transmil}. The experiments section of the manuscript reveals a notable performance decline in most MIL methods. 

\subsection{Vision Transformer ImageNet Pre-Trained}
We employ the same threshold filter technique for patch extraction as we have done in the ResNet18 scheme. Each patch is transformed into a 768-dimensional vector using a vision transformer pre-trained on ImageNet. The middle layer output dimension in MIL methods with feed-forward layers, such as TransMIL~\cite{shao2021transmil}, ILRA~\cite{lowrankmil}, DTFD~\cite{zhang2022dtfd}, ABMIL~\cite{ilse2018attention}, and the proposed method, is set to 512. In line with the TransMIL study, the output dimension of its middle layers is also established at 512.

\subsection{Warm-up Training Strategy}
\label{warmuptrainingstrategy}
As outlined in our paper, a warm-up training strategy is incorporated in all experiments of the proposed method. This warm-up training can be described as follows:

\begin{equation}
    \mathcal{L}_{final} = \begin{cases} \mathcal{L}_{ce} , \ \text{iff} \ t < 20, \\ \mathcal{L}_{ce} + \lambda_{tri} \mathcal{L}_{tri} + \lambda_{div}\mathcal{L}_{div}, \ \text{iff} \ t > 20, \end{cases}
    \label{eqn:1}
\end{equation}
where t is the current epoch. The total training epoch is set to 200  for all experiments in this paper. We only employ the cross entropy classification loss to train our model at the first 20 epochs; while adding all the other losses for the latter epochs. The rationale behind this is that the randomly initialized global vectors usually lead to instability in training. The warmup training will help the global vectors to learn the meaningful instance relation in classification. This prevents poorly initialized global vectors from incorrectly misleading the modeling of instance correlations at the start.

\begin{table*}[t]
\centering
\caption{Comparision over efficiency among different transformer-based MIL aggregation methods in terms of the number of Parameters (M) and MACs (G) represent the model size and multiple-accumulated operation computational complexity, respectively.}
\label{computercomplexity}
\resizebox{0.5\columnwidth}{!}{
\begin{tabular}{l|cc}
\toprule %
     Models & Params(M) & MACs(G) \\
    \cmidrule(r){1-3}
    ILRA-MIL~\cite{lowrankmil}&  1.049  & 1.842\\
    Trans-MIL~\cite{shao2021transmil}& 3.040 & 2.409\\
    Our & 0.642 & 1.054\\

\bottomrule
\end{tabular}
}
\end{table*}

\section{Efficiency Comparision of Transformer-Based MIL Aggregation Methods}
Take feature vectors extracted by ResNet-18 as an example, we apply the same hidden parameters as reported in the experiments. As shown in Table~\ref{computercomplexity}, the proposed method demonstrates superior efficiency compared to the other two transformer-based MIL aggregation methods, exhibiting notable advantages in terms of both model size and computational complexity. The cross-attention mechanism is more computationally efficient compared to the self-attention mechanism used across all instances. This efficiency stems from the use of an extremely short sequence of global vectors, which is substantially less in number than the total count of instances.

\noindent It is worth noting that ILRA-MIL~\cite{lowrankmil} employs self-attention for modeling the correlation between instances. Similarly, it also presents a larger number of parameters and is more computationally complex than the proposed method. The main reason is that they rewrite self-attention instead of self-attention with linear, and added the non-local pooling extra module. complexity~\cite{wang2020linformer,shen2021efficient,guo2022beyond}.

\section{Complexity of Diversity Loss}
The proposed diversity loss can be computed in a linear time complexity. 
For a global vector $\boldsymbol{G}\in \mathbb{R}^{K \times L}$, $\log\operatorname{det}(\boldsymbol{G}\boldsymbol{G}^T)=\sum_{i=1}^K\log(\lambda_i^2)$, where the main overhead is an SVD decomposition of $\boldsymbol{G}$ to get $\lambda_i$, resulting in a complexity of $\mathcal{O}(LK^2)\approx \mathcal{O}(L) $ due to $K$ is often set a small number (e.g., 5).

\section{Statistical Test}
\begin{figure}[h]
    \centering
    \includegraphics[width=0.7\columnwidth]{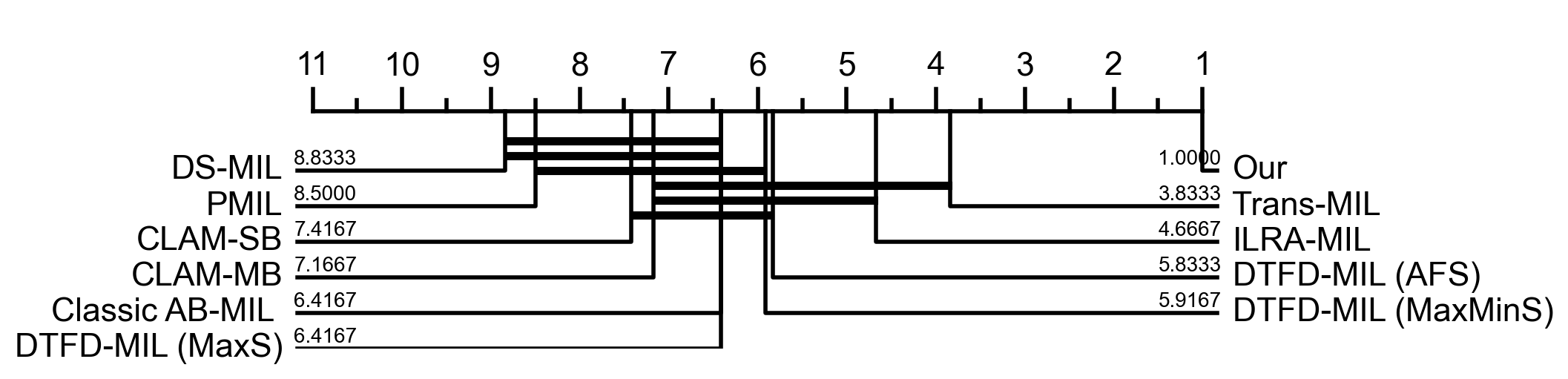}
    \vspace{-0.3cm}
    \caption{Wilcoxon signed-rank test, average rank denoted by the number. No statistical difference found between methods connected with one thickness line in the critical difference diagram.
}
    \label{fig:cd}
    \vspace{-0.3cm}
\end{figure}

 We present the Wilcoxon signed-rank test and the critical difference diagram \cite{demvsar2006statistical} in Fig. \ref{fig:cd} with $\alpha=0.5$ significance level. Our method statistically outperforms all other competitors.

\section{Additional Results}
We present the additional experiments on CAMELYON16 with CTransPath feature extractor and classic MIL benchmarks. The results are shown in Table \ref{tab:ctrans} and Table \ref{tab:classic}, respectively.

\begin{table}[h]
\caption{Results on \textbf{CTransPath} extractor. We employ the 4-fold cross-validation using data split provided by DTFD.}
\label{tab:ctrans}
		\centering
		\resizebox{0.9\textwidth}{!}{
			\begin{tabular}{lccc}
				\toprule
                     & \multicolumn{3}{c}{CAMELYON16} \\
                      \cmidrule(r){2-4}
				  & Accuracy & F1 &  AUC  \\
				\midrule
                    Classic AB-MIL (\textit{ICML'18})   & $0.940_{(0.933,0.948)}$& $0.936_{(0.928,0.944)}$ & $0.951_{(0.932,0.970)}$\\ %
   
                    DS-MIL (\textit{CVPR'21})  &$0.929_{(0.898,0.959)}$  &$0.923_{(0.889,0.957)}$  &$0.942_{(0.916,0.968)}$  \\ %
                    
                    Trans-MIL (\textit{NeurIPS'21}) & $0.952_{(0.935,0.970)}$  & $0.949_{(0.930,0.968)}$  & $0.973_{(0.958,0.987)}$ \\ %

                    DTFD-MIL (MaxMinS) (\textit{CVPR'22})  &$0.949_{(0.931,0.953)}$ & $0.933_{(0.906,0.937)}$ & $0.985_{(0.976,0.994)}$ \\ %
                    DTFD-MIL (AFS) (\textit{CVPR'22})  & $0.942_{(0.931,0.953)}$ & $0.922_{(0.906,0.937)}$ & $0.982_{(0.969,0.995)}$
                    \\%
                    ILRA-MIL (\textit{ICLR'23}) 
                    &$0.940_{(0.924,0.957)}$& $0.937_{(0.922,0.953)}$ & $0.961_{(0.946,0.975)}$ 
                      \\                 
                     \rowcolor{blue!8} 
                     \textbf{Our}  
&$\mathbf{0.972}_{(0.965, 0.979)}$& $\mathbf{0.971}_{(0.963, 0.978)}$ & $\mathbf{0.994}_{(0.991, 0.996)}$ \\ 
				\bottomrule
 			\end{tabular}}
    \vspace{-0.2cm}
\end{table}

\begin{table}[h]
  \caption{Results on MIL benchmarks.}
  \centering
  \resizebox{0.9\textwidth}{!}{
  \begin{tabular}{l|ccccc}
    \toprule
    Methods     & MUSK1  & MUSK2 & FOX & TIGER & ELEPHANT \\
    \hline
     mi-Net & 0.889 $\pm$ 0.039  & 0.858 $\pm$ 0.049 & 0.613 $\pm$ 0.035 &  0.824 $\pm$ 0.034 &  0.858 $\pm$ 0.037   \\
    MI-Net &  0.887 $\pm$ 0.041 & 0.859 $\pm$ 0.046 & 0.622 $\pm$ 0.038 & 0.830 $\pm$ 0.032 & 0.862 $\pm$ 0.034       \\
    MI-Net with DS & 0.894 $\pm$ 0.042 & 0.874 $\pm$ 0.043 & 0.630 $\pm$ 0.037 & 0.845 $\pm$ 0.039 & 0.872 $\pm$ 0.032    \\
    MI-Net with RC & 0.898 $\pm$ 0.043 & 0.873 $\pm$ 0.044 & 0.619 $\pm$ 0.047 & 0.836 $\pm$ 0.037 & 0.857 $\pm$ 0.040    \\
    ABMIL &  0.892 $\pm$ 0.040 & 0.858 $\pm$ 0.048 & 0.615 $\pm$ 0.043 & 0.839 $\pm$ 0.022 & 0.868 $\pm$ 0.022   \\
    ABMIL-Gated & 0.900 $\pm$ 0.050 & 0.863 $\pm$ 0.042 & 0.603 $\pm$ 0.029 & 0.845 $\pm$ 0.018 & 0.857 $\pm$ 0.027   \\
    DP-MINN  & 0.907 $\pm$ 0.036 & 0.926 $\pm$ 0.043 & 0.655 $\pm$ 0.052 & 0.897 $\pm$ 0.028 & 0.894 $\pm$ 0.030  \\
    NLMIL & 0.921 $\pm$ 0.017 & 0.910 $\pm$ 0.009 & 0.703 $\pm$ 0.035 & 0.857 $\pm$ 0.013 & 0.876 $\pm$ 0.011 \\
    ANLMIL & 0.912 $\pm$ 0.009 & 0.822 $\pm$ 0.084 & 0.643 $\pm$ 0.012 & 0.733 $\pm$ 0.068 & 0.883 $\pm$ 0.014 \\
    DSMIL & 0.932 $\pm$ 0.023  & 0.930 $\pm$ 0.020 & 0.729 $\pm$ 0.018 & 0.869 $\pm$ 0.008 & 0.925 $\pm$ 0.007 \\
    \hline
    Our Method & \textbf{0.989 $\pm$ 0.033} & \textbf{0.970 $\pm$ 0.0458} & \textbf{0.785 $\pm$ 0.120} & \textbf{0.925 $\pm$ 0.055}  & \textbf{0.950 $\pm$ 0.044}   \\
    \bottomrule
  \end{tabular}
  }
  \label{tab:classic}
  \vspace{-0.25cm}
\end{table}

